\pdfoutput=1
\documentclass[11pt,twoside,a4paper,cmspaper,final,collab]{cms-tdr}
\def\svnVersion{527156b}\def\svnDate{2026/10/23}\def\cmsCernNoTag{CERN-EP-2026-130}\def\cmsCernDate{\today}\def\cmsMessage{Published in the Journal of High Energy Physics as \href{http://dx.doi.org/10.1007/JHEP10(2026)186}{\doi{10.1007/JHEP10(2026)186}.}}
\begin{document}\cmsNoteHeader{SUS-21-005}

\newcommand{\mll}{\ensuremath{m_{\Pell\Pell}}\xspace}
\newcommand{\MJ}{\ensuremath{M_{\mathrm{J}}}\xspace}
\newcommand{\Nb}{\ensuremath{N_{\PQb}}\xspace}
\newcommand{\wjets}{\PW{}+jets\xspace}
\newcommand{\dyjets}{DY+jets\xspace}
\newcommand{\Nlep}{\ensuremath{N_{\text{lep}}}\xspace}
\newcommand{\Njet}{\ensuremath{N_{\text{jet}}}\xspace}

\newlength\cmsTabSkip\setlength{\cmsTabSkip}{1ex}
\providecommand{\cmsTable}[1]{\resizebox{\textwidth}{!}{#1}}

\cmsNoteHeader{SUS-21-005}
\title{Search for new physics using single-lepton events with high multiplicities of jets and \texorpdfstring{\PQb}{b} jets in proton-proton collisions at \texorpdfstring{$\sqrt{s}=13\TeV$}{sqrt(s)=13 TeV}}

\date{\today}

\abstract{
This paper presents a search for beyond the standard model physics using single-lepton events with a high multiplicity of jets, including those identified as bottom quark jets, without a requirement on missing transverse momentum. The analysis is based on proton-proton collision data collected with the CMS detector at the CERN LHC at a center-of-mass energy of 13\TeV, corresponding to an integrated luminosity of 138\fbinv. This search is sensitive to $R$-parity violating supersymmetry models, where supersymmetric particles can decay into standard-model particles through interactions that violate baryon number conservation. In particular, the signal model considered is gluino pair production, where each gluino decays into top, bottom, and strange quarks. The sum of large-radius jet masses is used to distinguish the signal from background, as it effectively captures the features of high jet multiplicity and high interaction energy. No significant excess of data over the background predictions is observed. Gluinos in this model have been excluded for masses below 1890\GeV at 95\% confidence level. 
}

\hypersetup{
pdfauthor={CMS Collaboration},%
pdftitle={Search for new physics using single-lepton events with high multiplicities of jets and b jets in proton-proton collisions at sqrt(s)=13 TeV},
pdfsubject={CMS},
pdfkeywords={CMS, supersymmetry, R-parity violation} 
}

\maketitle 
\section{Introduction}
\label{sec:intro}

The standard model (SM) has been remarkably successful in describing a wide array of phenomena, but it remains incomplete as a comprehensive framework for fundamental particles and their interactions. There are still theoretical challenges that remain unanswered, such as explaining the observed particle mass hierarchies in the presence of quantum corrections~\cite{BARBIERI198863,tHooft:1979rat,WITTEN1981513,DINE1981575,DIMOPOULOS1981353,DIMOPOULOS1981150,KAUL198236,NILLES19841,HABER198575}. In addition, the running of the gauge couplings in the SM does not lead to a unification at a single energy scale~\cite{PhysRevD.24.1681,IBANEZ1981439,EINHORN1982475,PhysRevD.25.3092}. Supersymmetry (SUSY) is a framework that offers a promising solution to these challenges.

In various SUSY models, a conserved quantum number called $R$-parity is introduced~\cite{FARRAR1978575} and defined as $(-1)^{2S+3(B-L)}$, where $S$ represents the particle's spin, while $B$ and $L$ denote its baryon and lepton numbers, respectively. The $R$-parity has significant implications for SUSY phenomenology. If $R$-parity is conserved, SUSY particles are produced in pairs at the CERN LHC, and their decay chains terminate in the lightest SUSY particle (LSP), which becomes a dark-matter candidate if it is electrically neutral. Since the LSP is stable in these models and interacts via the weak force, the presence of two neutral LSPs in the final state typically produces a signature characterized by large missing transverse momentum (\ptmiss). Searches targeting SUSY models that conserve $R$-parity generally require large \ptmiss. They have limited sensitivity to the $R$-parity violating (RPV) SUSY models because the LSPs are allowed to decay into SM particles, eliminating the \ptmiss signature. While RPV SUSY models do not contain the conventional stable LSPs, viable dark-matter candidates may still arise in extended or alternative sectors. In addition, they can address significant theoretical limitations of the SM~\cite{BARBIER20051}. Exploring both scenarios is essential for a comprehensive examination of the SUSY parameter space.

If $R$-parity violation is permitted, the following additional term ($W$) in the superpotential is allowed~\cite{BARBIER20051}:
\begin{equation*}
\label{eq:RPVLagrangian}
W = \frac{1}{2}\lambda^{ijk}L_{i}L_{j}\bar{e}_{k}+\lambda^{\prime ijk}L_{i}Q_{j}\bar{d}_{k}+\mu^{\prime i}L_{i}H_{u}
+\frac{1}{2}\lambda^{\prime\prime ijk}\bar{u}_{i}\bar{d}_{j}\bar{d}_{k}.
\end{equation*}
The fields $L_{i}$, $Q_{j}$, and $H_u$ are SU(2) doublets corresponding to leptons, quarks, and the Higgs boson, while the singlet fields $\bar{e}_k$, $\bar{u}_i$, and $\bar{d}_j$ represent the charged leptons, up- and down-type quarks, respectively. Color indices are omitted, and the indices $i,j,k$ denote the three different generations. The parameters $\lambda$ and $\mu$ stand for the coupling strengths associated with the corresponding interactions.

\begin{figure}[htbp]
  \centering
  \includegraphics[width=0.4\textwidth]{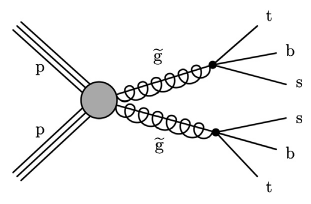}
	\caption{Diagram of the signal model considered in this analysis. Gluinos are produced in pairs and decay to \PQt, \PQb, and \PQs.}
  \label{fig:fd} 
\end{figure}

The search described in this paper is sensitive to signals with large multiplicities of jets and bottom quark (\PQb) jets, and low \ptmiss in single-lepton final states. A specific model of RPV within SUSY is used for the interpretation of the results, known as minimal flavor-violating SUSY~\cite{ref:MFV}. In this framework, the RPV couplings are derived from the SM Yukawa couplings, and this causes the third-generation RPV Yukawa couplings to dominate over those for the first two generations. This hierarchy is consistent with the stringent experimental constraint from the lack of neutron-antineutron oscillations~\cite{PhysRevD.103.012008}, which imposes strong limits on baryon number violation involving the lighter two generations. In particular, we are interested in the production of gluino pairs where the gluino decays to a top quark (\PQt) and a top squark (\PSQt). The \PSQt subsequently decays into \PQb and strange quark (\PQs) via the $\lambda^{\prime\prime}_{323}$ coupling ($\PSg\to\PQt\PASQt\to\PQt\PQb\PQs$). For the simplified model interpretation, several assumptions are made following Ref.~\cite{2018114}, including that \PSQt is virtual. In this scenario, the gluino decays directly to \PQt, \PQb, and \PQs, and the branching fraction for $\PSg\to\PQt\PQb\PQs$ is 100\%. The diagram of this model is shown in Fig.~\ref{fig:fd}.

There have been several searches for this model with the CMS experiment. Reference~\cite{CMS:2016zgb} reports searches in the 0, 1, 2, and ${\ge}3$ charged-lepton modes at $\sqrt{s} = 8\TeV$. Reference~\cite{2018114} reports the results with the 2016 data at 13\TeV in the 1-lepton final state. The ATLAS experiment has reported results using the full Run~2 data set, excluding gluino masses up to 1830\GeV at 95\% confidence level (\CL)~\cite{ATLAS:2021fbt}. Additional ATLAS searches have targeted gluino decays through the $\lambda^{\prime\prime}_{313}$ coupling, resulting in the \PQt, \PQb, and down quark final state~\cite{ATLAS:2023afl}, which is experimentally indistinguishable from the $\PQt\PQb\PQs$ final state considered in this analysis. ATLAS has also explored a broader set of RPV SUSY models, with exclusion limits presented as a function of $\lambda^{\prime\prime}_{323}$ and the gluino mass~\cite{ATLAS:2026mla}.

We search for beyond-the-SM physics in regions determined by the jet multiplicity, \Njet, and the number of identified bottom quark (\PQb-tagged) jets, \Nb. In each region, a signal would appear as an excess at high values of the sum of the masses of large-radius jets, \MJ. To suppress background and maintain a high signal acceptance, we target signal events where one of the top quarks decays leptonically. Therefore, the presence of an identified electron or muon is required. Hereafter, \Nlep denotes the number of identified electrons or muons. The quantity \MJ, originally proposed in phenomenological studies~\cite{Hook:2012fd,Cohen:2012yc,Hedri:2013pvl}, was used in searches by ATLAS and CMS for the all-hadronic~\cite{Aad:2015lea,Aad:2013wta} and single-lepton final states~\cite{Khachatryan:2016uwr,PhysRevLett.119.151802,2018114}, respectively. It captures the characteristics of high jet multiplicity and large interaction energy, thereby providing discrimination between signal and backgrounds. After applying corrections derived from comparisons of the \MJ templates between data and simulation in control regions, the distribution of \MJ is found to be reliably modeled. However, the \MJ distribution can have residual dependence on the event kinematics, particularly \Njet in the event. Thus, to avoid relying on simulation to accurately model the tails of the \Njet and \Nb distributions, the background estimation is performed using data-driven techniques in bins of \Njet and \Nb. By binning in \Nb and normalizing backgrounds separately in each \Nb bin, the analysis becomes less sensitive to any mismodeling of the \Nb shape, such as that arising from the treatment of high-momentum gluons splitting into two \PQb quarks.

The remainder of the paper is organized as follows. Section~\ref{sec:cms} describes the CMS detector, data, and simulated samples. The object reconstruction and event selections are presented in Section~\ref{sec:selection}. Section~\ref{sec:strategy} details the analysis strategy and background estimation, followed by the systematic uncertainties in Section~\ref{sec:syst}. The results and their interpretation are presented in Section~\ref{sec:results}, and the paper concludes with a summary in Section~\ref{sec:summary}. Tabulated results are provided in the HEPData record for this analysis~\cite{hepdata}.

\section{The CMS detector, data, and simulated samples}
\label{sec:cms}

This analysis uses the complete sample of proton-proton ($\Pp\Pp$) collisions at a center-of-mass energy of $\sqrt{s} = 13\TeV$ corresponding to an integrated luminosity of 138\fbinv~\cite{lumi_16,CMS-PAS-LUM-17-001,CMS-PAS-LUM-18-002}, collected with the CMS detector during 2016--2018.

The CMS apparatus~\cite{CMS:2008xjf,CMS:2023gfb} is a multipurpose, nearly hermetic detector, designed to trigger on~\cite{CMS:2020cmk,CMS:2016ngn,CMS:2024aqx} and identify electrons, muons, photons, and (charged and neutral) hadrons~\cite{CMS:2020uim,CMS:2018rym,CMS:2014pgm}. Its central feature is a superconducting solenoid of 6\unit{m} internal diameter, providing a magnetic field of 3.8\unit{T}. Within the solenoid volume are a silicon pixel and strip tracker, a lead tungstate crystal electromagnetic calorimeter (ECAL), and a brass and scintillator hadron calorimeter, each composed of a barrel and two endcap sections. Forward calorimeters extend the pseudorapidity ($\eta$) coverage provided by the barrel and endcap detectors. Muons are reconstructed using gas-ionization detectors embedded in the steel flux-return yoke outside the solenoid. More detailed descriptions of the CMS detector, together with a definition of the coordinate system used and the relevant kinematic variables, can be found in Refs.~\cite{CMS:2008xjf,CMS:2023gfb}.

In this search, Monte Carlo (MC) simulation samples are used for the background prediction. Samples for top quark pair production (\ttbar), a \PW boson produced with additional jets (\wjets), Drell--Yan production with additional jets (\dyjets), and multiple jets due to quantum chromodynamics interactions (QCD multijet) are generated using \MGvATNLO~\cite{Alwall:2014hca} v2.6.5 and v2.7.3 at leading-order (LO) accuracy, with 3, 4, 4, and 2 extra partons in the matrix-element calculations, respectively. Samples for top quark pair production in association with a \PW boson ($\ttbar\PW$), \PZ boson ($\ttbar\PZ$), and Higgs boson ($\ttbar\PH$), as well as for four top quark production ($\ttbar\ttbar$) and $s$-channel single top quark production are generated by \MGvATNLO at next-to-LO (NLO) accuracy. The single top quark production with a \PW boson and $t$-channel single top quark process are generated with {\POWHEG} v2.0~\cite{Nason:2004rx,Frixione:2007vw,Alioli:2010xd}, while the diboson and triboson processes are generated with {\PYTHIA} 8.240~\cite{pythia8}. The signal samples are generated by \MGvATNLO v2.6.5 at LO accuracy, using MSSM\_SLHA2 Universal \textsc{FeynRules} Output model~\cite{ref:MSSM_SLHA2,ref:UFO}, with two extra partons in the matrix-element calculations. The signal production cross sections are approximated to next-to-NLO (NNLO) plus next-to-next-to-leading logarithmic (NNLL) accuracy~\cite{10.21468/SciPostPhysCore.7.4.072}. The signal sample assumes that the gluino is the LSP and that all other particles are decoupled by taking their masses sufficiently large that they do not affect the phenomenology. All samples are generated with the NNPDF3.1 set of parton distribution functions~\cite{ref:NNPDF31} at NNLO accuracy. Samples generated at NLO accuracy with the \MGvATNLO use the FxFx matching and merging scheme~\cite{FxFx}, while those generated at LO accuracy with the same generator employ the MLM merging scheme~\cite{MLM}. Fragmentation and hadronization processes are simulated by \PYTHIA, and underlying events are described by CP5 tune~\cite{CMS:2019csb}. Finally, detector response is simulated by \GEANTfour~\cite{Agostinelli:2002hh}.     

\section{Object and event selection} 
\label{sec:selection}

The reconstruction of physics objects in an event proceeds from the candidate particles identified by the particle-flow (PF) algorithm~\cite{Sirunyan_2017}, which uses information from the tracker, calorimeters, and muon systems to identify the candidates as charged or neutral hadrons, photons, electrons, or muons. Charged-particle tracks are required to originate from the primary vertex (PV) of the event. It is taken as the vertex corresponding to the hardest scattering in the event, evaluated using tracking information alone, as described in Ref.~\cite{CMS:2014pgm}.

The charged PF candidates associated with the PV and the neutral PF candidates are clustered into jets using the anti-\kt algorithm~\cite{Cacciari:2008gp} with a distance parameter $R=0.4$, as implemented in the \FASTJET package~\cite{Cacciari:2011ma}. The estimated pileup contribution to the transverse momentum (\pt) of a jet from neutral PF candidates is removed with a correction based on the area of the jet and the average energy density of the event~\cite{Cacciari:2007fd}. The jet energy is calibrated using $\pt$- and $\eta$-dependent corrections; the resulting calibrated jet is required to satisfy $\pt>30\GeV$ and $\abs{\eta}<2.4$. Each jet must also meet the tight identification requirements specified in Ref.~\cite{CMS-PAS-JME-16-003} to suppress, for example, calorimeter noise. Finally, as defined below, jets with PF constituents matched to the selected lepton are removed from the jet collection.

A subset of the jets is ``tagged'' as originating from \PQb quarks using the \textsc{DeepCSV} algorithm~\cite{Sirunyan:2017ezt}, with performance evaluated using simulated \ttbar events. The signal efficiency for \PQb jets at the medium working point (WP) is 68\%. The probability of misidentifying jets arising from \PQc quarks is 12\%, while from light-flavor quarks or gluons is 1.1\%. 

Electrons are reconstructed by associating a charged-particle track with an ECAL supercluster. The resulting candidate electrons are required to satisfy $\pt>20\GeV$ and $\abs{\eta}<2.5$, excluding the barrel-endcap transition region since the reconstruction of an electron object in this region is not optimal. Additionally, they must pass the medium WP of the identification criteria based on a multivariate analysis~\cite{CMS:2020uim}. Muons are reconstructed by associating tracks in the muon system with those found in the silicon tracker. Muon candidates are required to have $\pt>20\GeV$ and $\abs{\eta}<2.4$ and to pass the medium cutoff-based identification criteria~\cite{CMS:2018rym}. Both types of candidate leptons are required to be associated with the PV. To select leptons from the \PW boson in the top quark decay, electrons and muons are required to have relative mini-isolation~\cite{miniso} less than 0.1 and 0.2, respectively. The relative mini-isolation is defined as the sum of the \pt of other PF candidates within a cone of radius $R_{\mathrm{mini}}$, divided by the lepton \pt, $\pt^{\Pell}$. $R_{\text{mini}}$ is given by 0.2 for $\pt^{\Pell}<50\GeV$, $10\GeV/\pt^{\Pell}$ for $50<\pt^{\Pell}<200\GeV$, and 0.05 for $\pt^{\Pell}>200\GeV$.

We use the anti-\kt algorithm to cluster $R=0.4$ jets and the selected isolated lepton into $R=1.2$ jets. Hereafter, the $R=0.4$ and $R=1.2$ jets are referred to as ``small-$R$'' and ``large-$R$'' jets, respectively. Clustering small-$R$ jets, rather than PF candidates, incorporates jet pileup corrections, thereby mitigating the mass dependence on pileup. The variable \MJ is defined as the sum of all large-$R$ jet masses, $\MJ=\sum_{J_i={\text{large-}R\text{ jet}}}m(J_i)$. This clustering technique, which combines small-$R$ jets into large-$R$ jets, has been previously employed by both CMS~\cite{2018114,khachatryan2016search} and ATLAS~\cite{Aad:2014bva}. Including leptons in the large-$R$ jets ensures the representation of the event's full kinematics. Throughout this analysis, quantities related to \Njet or \Nb are based solely on small-$R$ jets, excluding contributions from large-$R$ jets.

The key issue in the analysis is suppression and estimation of the dominant SM backgrounds, which arise primarily from the production of $\ttbar$, QCD multijet, and \wjets. Backgrounds from \dyjets, single top quark, multiboson, and $\ttbar+(\PW,\PZ,\PH,\ttbar)$ productions are also considered but are relatively small. Analysis regions are defined by requiring one isolated lepton, at least four small-$R$ jets, and $\HT>1200\GeV$, where $\HT$ is the scalar sum of the \pt of selected small-$R$ jets. We further suppress the backgrounds by requiring $\MJ>500\GeV$. The combination of these selections is referred to as the baseline selection.

The 2016 data sample used for this study is obtained using a set of triggers that require at least one jet with $\pt>450\GeV$ and $\HT>900\GeV$. For the 2017--2018 samples, a trigger that requires $\HT>1050\GeV$ is used. The analysis targets events with large hadronic activity, corresponding to high \HT and high \MJ, for which these triggers are fully efficient after the baseline selection. The trigger efficiency is consistent with 100\% and does not depend on the analysis variables.

\section{Analysis strategy and background prediction} 
\label{sec:strategy}

\begin{figure}[htbp]
\centering
  \includegraphics[width=0.32\textwidth]{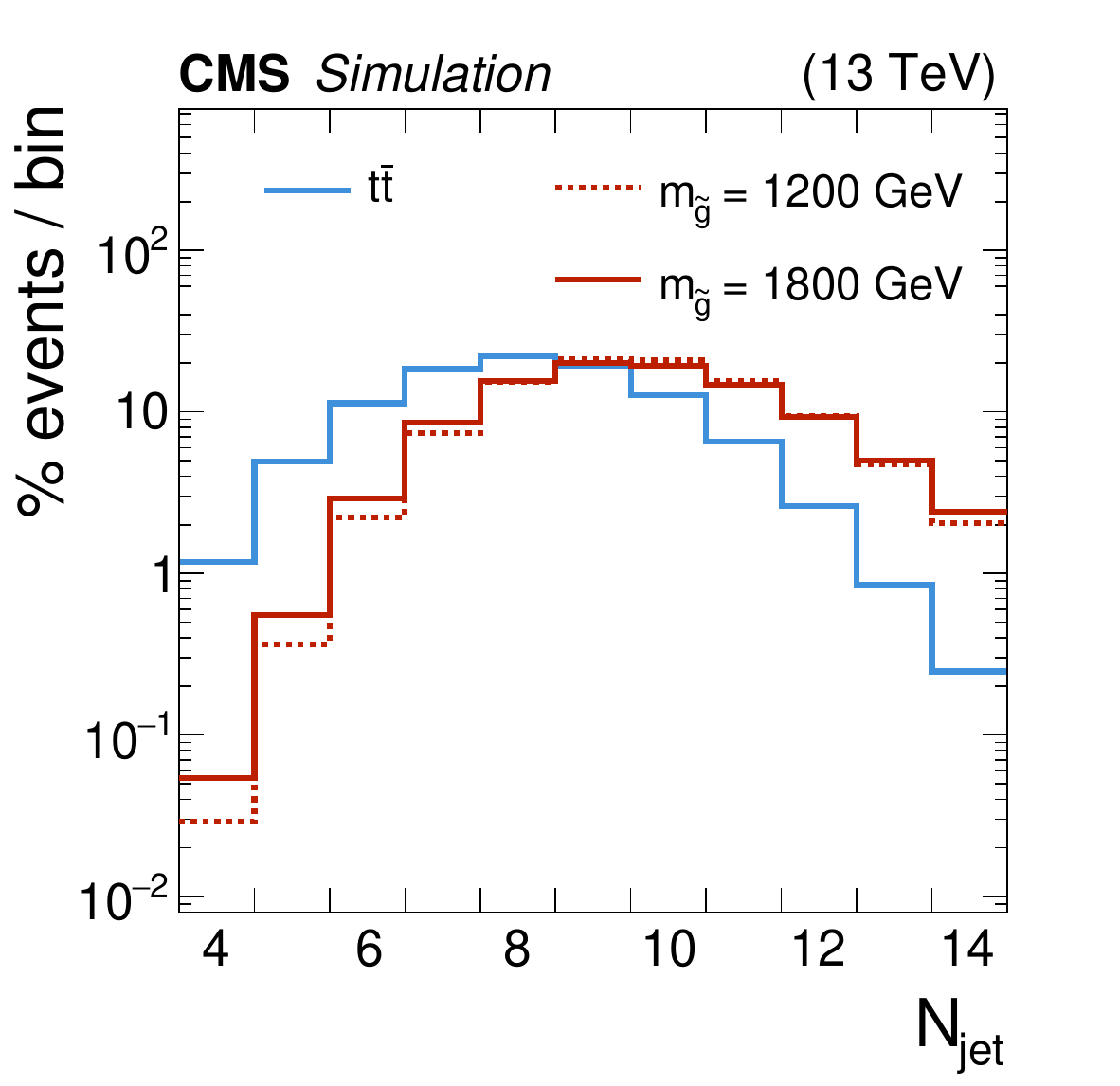}
  \includegraphics[width=0.32\textwidth]{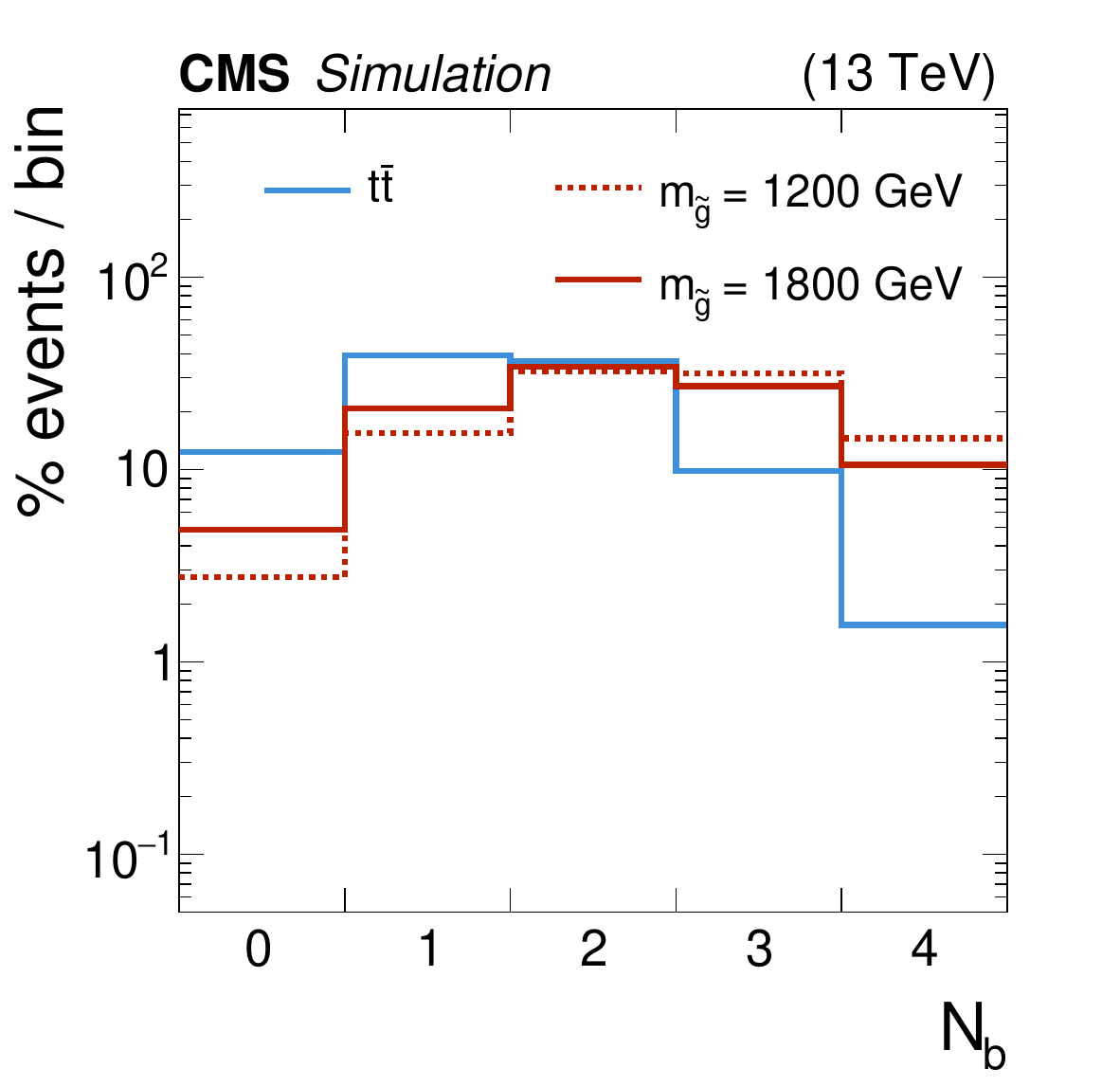}
  \includegraphics[width=0.32\textwidth]{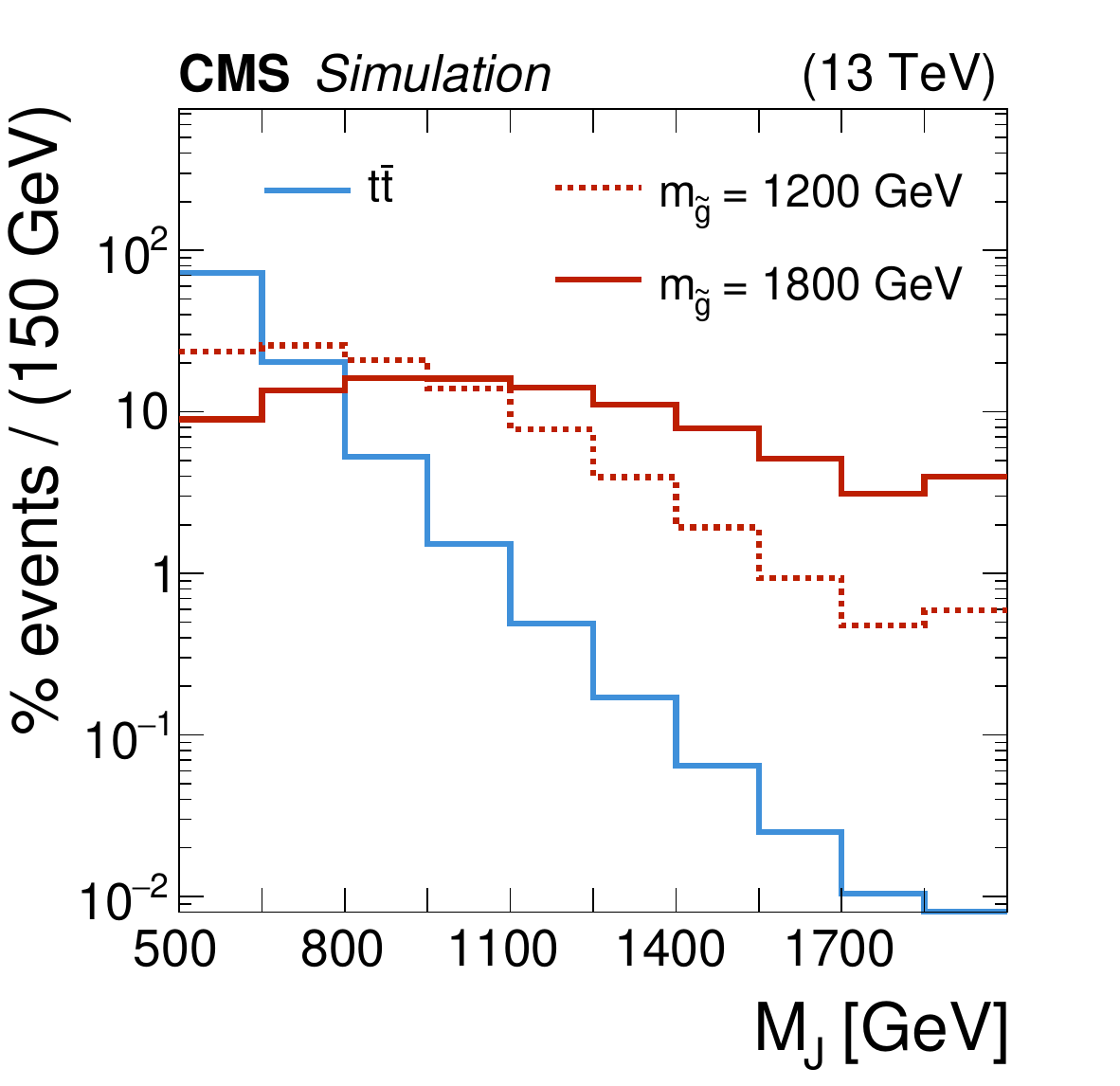}
  \caption{Distributions of \Njet (left), \Nb (middle), and \MJ (right) for $\ttbar$ and signal events in two gluino mass scenarios, after applying the baseline selection, as described in Section~\ref{sec:selection}.} 
\label{fig:var_1d}
\end{figure}

The distributions of \Njet, \Nb, and \MJ for $\ttbar$ and signal events after the baseline selection are shown in Fig.~\ref{fig:var_1d}. The signal process with one lepton in the final state contains at least eight quarks, including four \PQb quarks, leading to the signal events predominantly populating the tails of these distributions. Motivated by this, the analysis regions are defined using \Njet and \Nb, with regions of high \Njet and \Nb designated as signal regions (SRs), as summarized in Table~\ref{tab:analysis_regions}. Regions denoted as control regions (CRs) contribute minimally to sensitivity and are primarily used to constrain systematic uncertainties of the background in the SR. Specifically, the \Njet bins are defined as 4--5, 6--7, and ${\geq}8$, and the \Nb bins as 0, 1, 2, 3, and ${\geq}4$. The two highest \Nb bins in the $4\le\Njet\le5$ region are merged due to limited data sample size. The corresponding \Nb bins in $\Nlep=0$ region are merged as well for consistency. In each (\Nb, \Njet) bin, the \MJ template is used to predict the background and discriminate the signal from it. The dominant background in the high-\Nb bins is $\ttbar$, while \wjets and QCD multijet dominate the low-\Nb bins.

\begin{table}[htbp]
	\topcaption{\label{tab:analysis_regions} A diagram representing the analysis regions after the baseline selection. Here, ``CR'' denotes a control region and ``SR'' denotes a signal region. The \MJ template is used in the $\Nlep=1$ regions (darker yellow CR and SR), while it is not used in the $\Nlep=0$ regions.}
  \centering
  \begin{tabular}{ccccccc}
    \hline
    \multirow{3}{*}{$\Nb$} & \multicolumn{3}{c}{$\Nlep=0$} & \multicolumn{3}{c}{$\Nlep=1$} \\ \cline{2-7}
			   & \multicolumn{3}{c}{$\Njet$}  & \multicolumn{3}{c}{$\Njet$} \\ \cline{2-7}
			   & 6--7 & 8--9 & ${\geq}10$ & 4--5 & 6--7 & ${\geq}8$  \\ \hline
	  $\,\,0\,\,$    & \cellcolor{yellow!30}{CR} & \cellcolor{yellow!30}{CR} & \cellcolor{yellow!30}{CR} & \cellcolor{yellow}{CR} & \cellcolor{yellow}{CR} & \cellcolor{yellow}{CR} \\
		 $\,\,1\,\,$    & \cellcolor{yellow!30}{CR} & \cellcolor{yellow!30}{CR} & \cellcolor{yellow!30}{CR} & \cellcolor{yellow}{CR} & \cellcolor{yellow}{CR} & \cellcolor{yellow}{CR} \\
	  $\,\,2\,\,$    & \cellcolor{yellow!30}{CR} & \cellcolor{yellow!30}{CR} & \cellcolor{yellow!30}{CR} & \cellcolor{yellow}{CR} & \cellcolor{yellow}{CR} & \cellcolor{pink}{SR} \\
	  $\,\,3\,\,$    & \cellcolor{yellow!30}{} & \cellcolor{yellow!30}{CR} & \cellcolor{yellow!30}{CR} & \cellcolor{yellow}{} & \cellcolor{pink}{SR} & \cellcolor{pink}{SR} \\
	  ${\ge}4$         & \multirow{-2}{*}{\cellcolor{yellow!30}{CR}} & \cellcolor{yellow!30}{CR} & \cellcolor{yellow!30}{CR} & \multirow{-2}{*}{\cellcolor{yellow}{CR}}   & \cellcolor{pink}{SR} & \cellcolor{pink}{SR} \\
    \hline
  \end{tabular}
\end{table}

In the $\Nlep=1$ analysis region, the \Njet and \Nb distributions may not be modeled perfectly by simulation. This is mitigated by binning the sample in \Njet and \Nb, and then fitting the \MJ~distribution to data. This approach allows the normalization of the main backgrounds---QCD multijet, \wjets, and $\ttbar$---to be determined in each (\Nb, \Njet) bin. The detailed fit model, including the treatment of the individual background normalizations and the associated systematic uncertainties, is described in Section~\ref{sec:results}. 

The previous study~\cite{2018114} used \Nb templates in bins of \Njet and \MJ, but this study uses \MJ templates in bins of \Njet and \Nb. By normalizing backgrounds separately in each \Nb bin, the analysis becomes less sensitive to systematic effects that can potentially alter the \Nb shape, such as gluon splitting and composition of jet flavor. Gluon-splitting effects were the main source of systematic uncertainty in the previous version of the analysis, and the new strategy helps to mitigate such effects.

The \MJ templates are taken from MC simulation with corrections applied for jet energy and \PQb tagging efficiency differences with respect to data. Even after applying these corrections, the modeling of \MJ shapes may still not be perfect and residual corrections are needed. We measure the correction factors, hereafter referred to as $\kappa$ factors, separately for each \Njet bin in independent control samples. The measured correction factors and their uncertainties are used as input to a global fit to the \MJ distributions in \Njet and \Nb bins, as discussed in Section~\ref{sec:results}. The \MJ templates have three bins: $\MJ=$ 500--800, 800--1100, and ${>}1100\GeV$. For each \MJ template, there are two $\kappa$ factors, $\kappa_1$ and $\kappa_2$. They are defined as
\begin{equation}
  \begin{aligned}
    \kappa_1 &= \frac{R_1^{\mathrm{data}}}{R_1^{\mathrm{MC}}}
              = \frac{\left( \frac{N(800<\MJ<1100\GeV)}{N(500<\MJ<800\GeV)} \right)_{\mathrm{data}}}
                     {\left( \frac{N(800<\MJ<1100\GeV)}{N(500<\MJ<800\GeV)} \right)_{\mathrm{MC}}}, \\
    \kappa_2 &= \frac{R_2^{\mathrm{data}}}{R_2^{\mathrm{MC}}}
              = \frac{\left( \frac{N(\MJ>1100\GeV)}{N(500<\MJ<800\GeV)} \right)_{\mathrm{data}}}
                     {\left( \frac{N(\MJ>1100\GeV)}{N(500<\MJ<800\GeV)} \right)_{\mathrm{MC}}}.
  \end{aligned}
\end{equation}
The \MJ shape is corrected using $\kappa$ factors. Residual dependencies of the \MJ shape on \Nb are addressed through systematic uncertainties, as described in Section~\ref{subsec:bkg_syst}. The fit model is constructed such that the $\kappa$ factors can be constrained by the large sample of events in the low-\Nb region. Our strategy is to derive conservative uncertainties in the $\kappa$ factors from the independent CRs and let the global fit further constrain the $\kappa$ factors, particularly using the low-\Nb region.        

The \MJ shapes of the main backgrounds are overall similar, but not identical. To address the possible dependence of $\kappa$ factors on processes, we use separate $\kappa$ factors for QCD multijet, \wjets, and $\ttbar$, and estimate them in dedicated measurement regions, as described in Table~\ref{tab:kappa_meas}. The regions for QCD multijet and \wjets are orthogonal to the $\Nlep=1$ analysis region. However, the region for the $\ttbar$ background has a partial overlap and the global fit does not use it to constrain the $\kappa$ factors. For QCD multijet, the measured $\kappa$ factors are used to correct the \MJ templates, while for $\ttbar$ and \wjets, they are treated as uncertainties and constrained by the global fit.

\begingroup
\renewcommand{\arraystretch}{1.4}
\begin{table}[htbp]
  \topcaption{Selection criteria applied for measuring the $\kappa$ factors of the main backgrounds.}
  \centering
\cmsTable{%
\begin{tabular}{lccccccc}
\hline
	Process & $\Nlep$ & \multicolumn{3}{c}{$\Njet$} & $\Nb$ & $\HT$ [GeV] & $\MJ$ [GeV] \\
\hline
	QCD multijet    & 0     & 5--6  & 7--8 & ${\ge}9$ & 0     & ${>}1200$ & ${>}500$ \\
	\wjets 	& \parbox[c]{4.5cm}{\centering Two opposite-sign, same-flavor leptons\\ with $81 < \mll < 101\GeV$}
	& 3--4 & 5--6 & ${\ge}7$ & ${\le}2$  & ${>}1200$ & ${>}500$ \\
	$\ttbar$     & 1     & 4--5 & 6--7 & ${\ge}8$ & 1 & ${>}1200$ & ${>}500$ \\
\hline
\end{tabular}
}
  \label{tab:kappa_meas}
\end{table}
\endgroup

The $\kappa$ factors for the QCD multijet process are measured in the $\Nlep=0$ and $\Nb=0$ region. After this selection, the purity of QCD multijet is close to 99\% in all \Njet bins. Though their contributions are small, other backgrounds are subtracted from data.

For \wjets, the $\kappa$ factors are measured using \dyjets events selected with two opposite-sign and same-flavor leptons (OSSF) whose invariant mass (\mll) satisfies $81 < \mll < 101\GeV$. The \Njet requirement in the selection is relaxed by $1$ because of the requirement $\Nlep=2$ (note that leptons are included in large-$R$ jet clustering). The purity of \dyjets events is higher than 90\% in all \Njet regions, and other contributions estimated from MC simulation are subtracted from data. Potential differences in \MJ shapes between \dyjets and \wjets backgrounds are estimated from simulation and incorporated as uncertainties in the $\kappa$ factors, showing agreement within 20\% in all \Njet bins.
 
For the $\ttbar$ process, the $\kappa$ factors are measured in the $\Nlep=1$, $\Nb=1$ region. This region is dominated by $\ttbar$ in the high-\Njet region, while \wjets dominates the low-\Njet region. The QCD multijet contribution is also non-negligible. Because the purity of $\ttbar$ events in this region is low, the measurement of $\ttbar$ $\kappa$ factors is prone to mismodeling of the \MJ shape of QCD multijet and \wjets backgrounds. To mitigate this, a dedicated fit---following the same procedure as the global fit but restricted to the $0\le\Nb\le1$ region---is performed to obtain more precise \MJ shapes for these backgrounds. In the $\Nb=1$ region, the fitted QCD multijet and \wjets contributions are then subtracted from data, and the $\ttbar$ $\kappa$ factors are derived.

\begin{figure}[htbp]
\centering
  \includegraphics[width=0.32\textwidth]{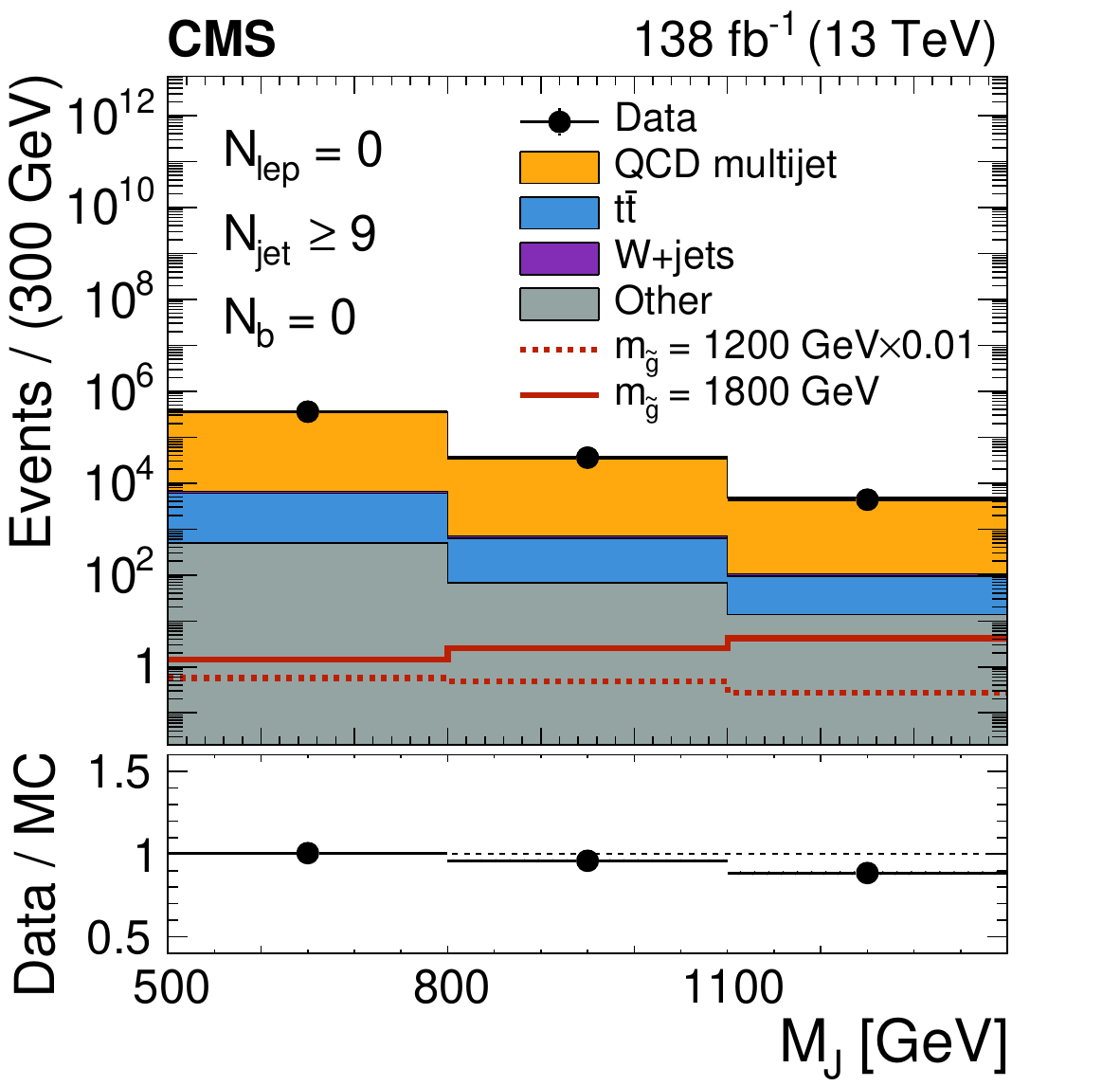}
  \includegraphics[width=0.32\textwidth]{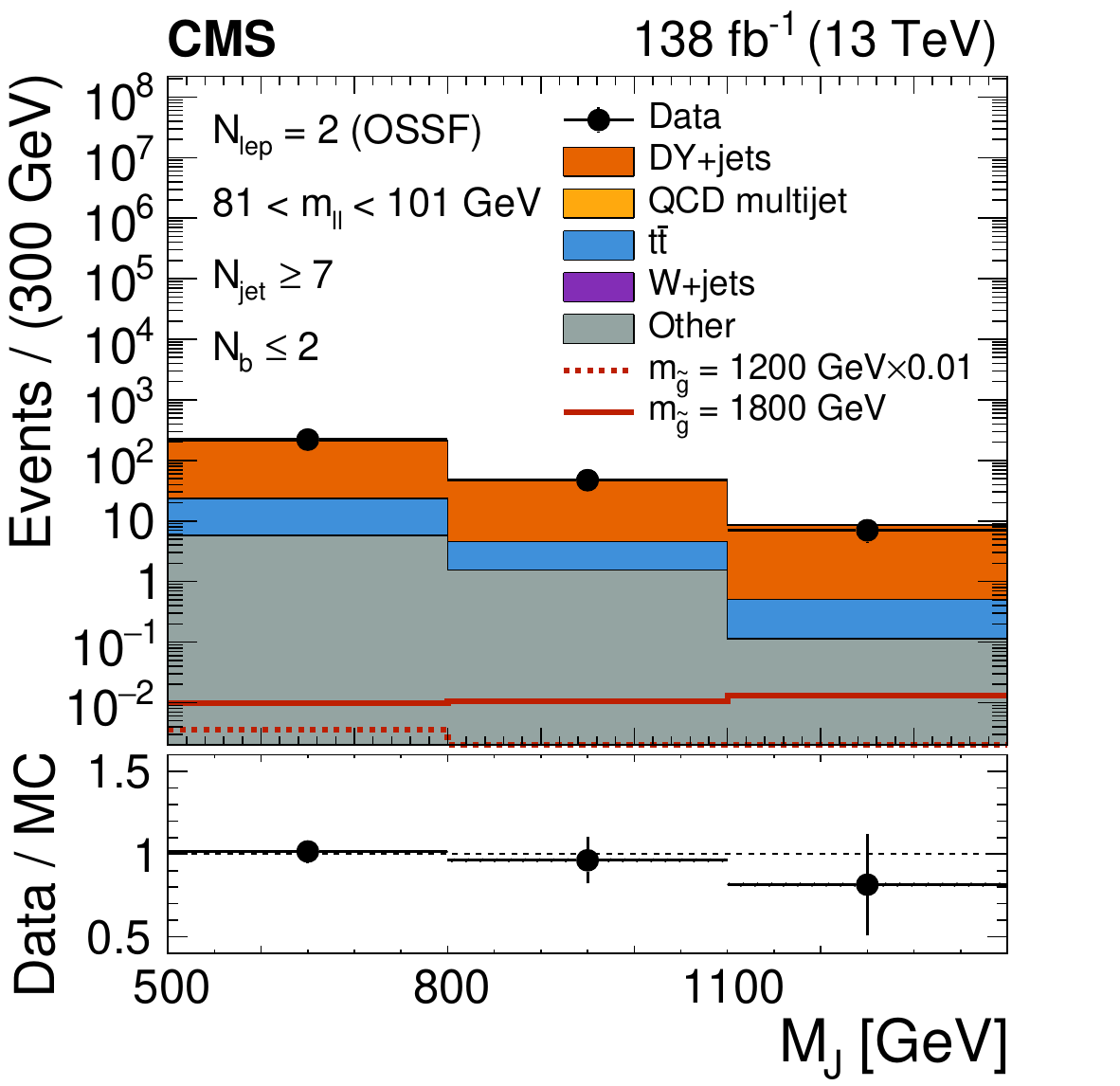}
  \includegraphics[width=0.32\textwidth]{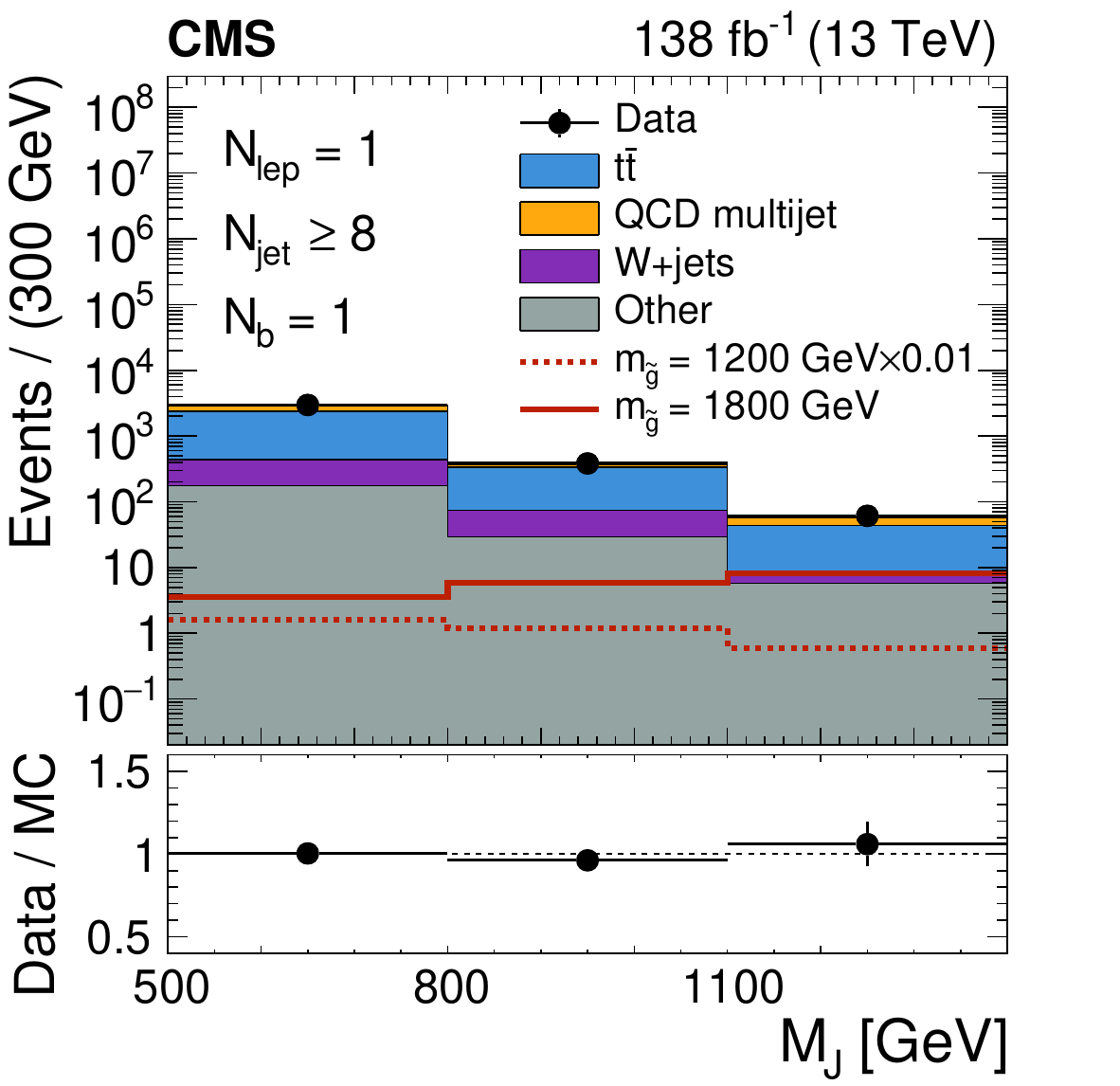}
	\caption{Comparison of \MJ shapes between data and simulation in the CRs used to measure the $\kappa$ factors for QCD multijet (left), \wjets (middle), and $\ttbar$ (right). The selection corresponds to $\Nlep=0$, $\Njet\ge9$, $\Nb=0$ for QCD multijet; a \dyjets background dominated region with $\Nlep=2$, $\Njet\ge7$, $\Nb\le2$ for \wjets; and $\Nlep=1$, $\Njet\ge8$, $\Nb=1$ for $\ttbar$. The simulation is normalized to data in all regions.}
\label{fig:kap_meas}
\end{figure}

Figure~\ref{fig:kap_meas} shows the \MJ shape comparisons between data and simulation in the $\kappa$ factor measurement regions for QCD multijet (left), \wjets (middle), and $\ttbar$ (right) in the high-\Njet region. The simulation is normalized to data. 

\begin{figure}[htbp]
\centering
  \includegraphics[width=0.55\textwidth]{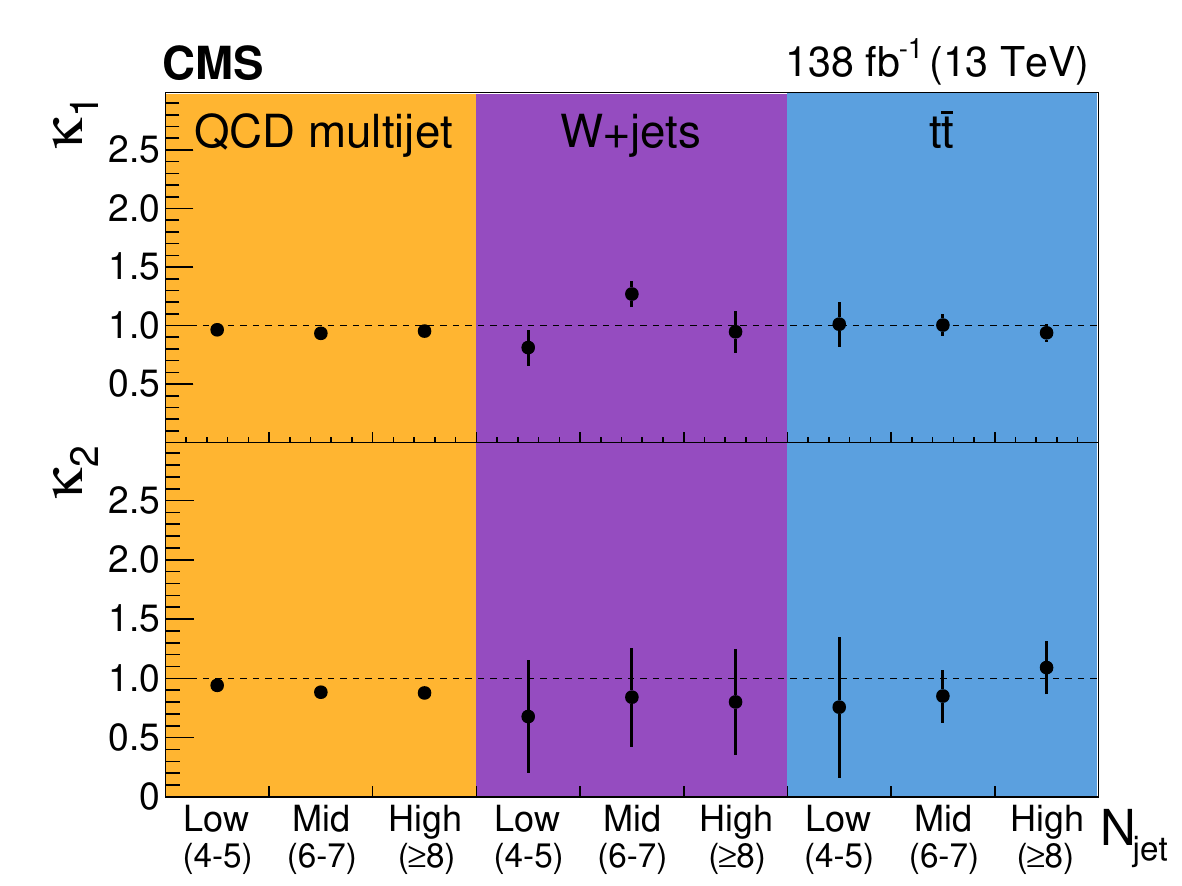}
\caption{The $\kappa$ factors measured in control regions for the QCD multijet (yellow left), \wjets (purple middle), and $\ttbar$ (blue right) backgrounds. The upper and lower panels correspond to $\kappa_1$ and $\kappa_2$, respectively. In each case, the three points indicate the $\kappa$ values for the \Njet bins used in the global fit. The error bars include the statistical uncertainties of data (dominant) and MC simulation samples.}
\label{fig:kappa_meas_summary}
\end{figure}

Figure~\ref{fig:kappa_meas_summary} shows the $\kappa$ factors measured in the dedicated CRs. Three colored areas show the measurements of QCD multijet (yellow), \wjets (purple), and $\ttbar$ (blue) $\kappa$ factors. In each region, the three measurements correspond to the three \Njet bins. All $\kappa$ factors for the QCD multijet background are measured with high precision, allowing us to use these $\kappa$ factors as corrections to the QCD multijet background \MJ shapes. Therefore, the QCD multijet background \MJ templates in $\Nlep=1$ region are corrected based on these $\kappa$ factors, and the difference from unity is used as the uncertainty.

On the other hand, $\kappa$ factors for $\ttbar$ and \wjets, particularly $\kappa_2$, have large uncertainties due to limited sample size in data. This makes the measurements prone to statistical fluctuations. Therefore, we do not correct the \MJ templates of $\ttbar$ and \wjets in the $\Nlep=1$ region but take the larger of the statistical uncertainty in the $\kappa$ factor and its deviation from unity as the shape variation uncertainty. Table~\ref{tab:kappa} shows the actual values of uncertainties of $\kappa$ factors used as input to the global fit. As mentioned already, the goal of the measurements in this section is to provide a reasonable initial range of $\kappa$ factors, which will then be constrained in the global fit.  

\begin{table}[htbp]
  \centering
  \topcaption{Uncertainties in the $\kappa$ factors for QCD multijet, \wjets, and $\ttbar$ backgrounds in the three $\Njet$ regions used in the global fit.}
  \begin{tabular}{llcc}
  \hline
  Process & \Njet & $\sigma_{\kappa_1}$[\%] & $\sigma_{\kappa_2}$[\%] \\
  \hline
  \multirow{3}{*}{QCD multijet}
  &4--5     & 4 & 6 \\
  &6--7     & 7 & 12 \\
  &${\ge}8$ & 5 & 12 \\[\cmsTabSkip]
  \multirow{3}{*}{\wjets}
  &4--5     & 19 & 48  \\
  &6--7     & 11 & 42  \\
  &${\ge}8$ & 18 & 44 \\[\cmsTabSkip]
  \multirow{3}{*}{$\ttbar$}
  &4--5     & 19 & 60  \\
  &6--7     & 9  & 22  \\
  &${\ge}8$ & 8  & 22 \\
  \hline
  \end{tabular}
  \label{tab:kappa}
\end{table}

\section{Systematic uncertainties} 
\label{sec:syst}

\subsection{Background systematic uncertainties}
\label{subsec:bkg_syst}

Since jet flavor information is not considered when constructing the \MJ variable, the \MJ distribution and the $\kappa$ factors are generally not expected to have a strong correlation with \Nb. Nevertheless, we estimate potential dependencies and incorporate them into the global fit. In particular, high \PQb jet multiplicity can lead to variations in the \MJ shape, which may arise from effects such as gluon splitting. In some \Nb regions, the \MJ shape was found to deviate from that of other \Nb bins beyond the statistical uncertainty of the MC simulation samples. This could be an issue because the $\kappa$ factors in the high-\Nb bins are constrained by those in the low-\Nb bins. To address this, we assign separate uncertainties in the \MJ shape to the primary background processes in each \Nb bin. For the \wjets background, the uncertainty is evaluated based on the \MJ shape difference between $\Nb=0$ and the other \Nb bins. For the $\ttbar$ background, uncertainties are derived from the difference of the \MJ shape between $\Nb=1$ and other \Nb bins, using the large number of events available in the $\Nb=1$ region. In the bin with $\Njet \ge 8$ and $\Nb \ge 4$, which drives the sensitivity of the analysis, the associated uncertainty ranges 3--10\% for $\kappa_1$, and 10--36\% for $\kappa_2$.

In addition, systematic effects can cause the \MJ templates to have different shapes across the \Nb bins. The impact of the jet energy scale (JES), jet energy resolution (JER), heavy-flavor jet \PQb tagging scale factors, light-flavor and gluon jet mistagging scale factors, renormalization and factorization scales, gluon-splitting description, muon efficiency, electron efficiency, and pileup is evaluated. For the gluon splitting uncertainty, the cross section of events in which gluons split into two quarks is varied by ${\pm}20$\% based on the measurements in Ref.~\cite{sirunyan2020measurement}. These variations are evaluated by modifying the \MJ templates according to different systematic sources. To quantify the impact of each systematic source, the $\kappa$ factors are recalculated using simulation samples of the main backgrounds where each systematic variation is applied. The resulting ${\pm}1$ standard deviation variations in the $\kappa$ factors are generally within the statistical uncertainty of the MC simulation samples, however, some systematic effects appear to show small but consistent trends. To address this, a linear fit of $\kappa$ factor as a function of \Nb is performed for each \Njet bin to estimate the potential \Nb-dependence of the $\kappa$ factor. This procedure is carried out separately for each systematic source and for each data-taking year. Most systematic sources contribute marginally; therefore, their effects, defined as $(\kappa - 1)$, are summed in quadrature within each \Nb bin. The JES and JER are excluded and treated separately due to their comparatively larger impacts. Finally, all contributions are incorporated as uncertainties in the $\kappa$ factors. For the $\ttbar$ background, in the bin with $\Njet\ge8$ and $\Nb\ge4$, the total uncertainty from these sources ranges 5--13\%.

\subsection{Signal systematic uncertainties}
\label{subsec:sig_syst}

Several systematic uncertainties affecting the signal yield are evaluated. The same systematic sources listed in Section~\ref{subsec:bkg_syst} are included. In addition, uncertainties associated with the trigger efficiency and initial-state radiation are considered. The trigger efficiency is consistent with 100\%. Based on their maximum variations in the region of $\HT>1200\GeV$, we assign a flat 1\% as the uncertainty in the trigger efficiency. Additionally, corrections for initial-state radiation are not applied, and a flat 1\% uncertainty is assigned based on the chosen simulation settings. The systematic uncertainties for a signal corresponding to $m_{\PSg}=1800\GeV$, evaluated over all three years in the $\Njet\ge8$ with $\Nb\ge4$ bin, are shown in Table~\ref{tab:signal_systs_bin36}. In the $\MJ>1100\GeV$ region of this bin, which provides the highest sensitivity, the dominant signal systematic uncertainties for all data-taking years arise from heavy-flavor jet \PQb tagging scale factors and the JES. 

\begin{table}[htbp]
  \centering
  \topcaption{List of signal systematic uncertainties and their effects on the yields in the $\Njet\ge8$ and $\Nb\ge4$ bin for all three years combined. The signal model corresponds to $m_{\PSg}=1800\GeV$.}
  \cmsTable{%
  \begin{tabular}{lrrr}
  \hline
	  \multirow{3}{*}{Systematic source} &  \multicolumn{3}{c}{Uncertainty [\%]} \\ \cline{2-4}
	  				     &  \multicolumn{3}{c}{\MJ [GeV]}  \\ 
					     &  500--800 & 800--1100 & ${>}1100$ \\ \hline
	  MC event count                                                    		 & 3.6        & 2.8         & 2.4       \\
	  Combination of renormalization and factorization scales          		 & 0.3        & 0.2         & 0.8       \\
	  Renormalization scale 				           		 & 0.4        & ${<}0.1$    & ${<}0.1$  \\
	  Factorization scale					           		 & 0.3        & 0.2         & 0.8       \\
	  Electron efficiency					           		 & 0.4        & 0.4         & 0.4       \\
	  Muon efficiency					           		 & ${<}0.1$   & ${<}0.1$    & ${<}0.1$  \\
	  Jet energy resolution					           		 & 0.7   & 0.5    & 0.5   \\
	  Jet energy scale					           		 & 0.5   & 1.0    & 3.6   \\
	  Across-year correlated \PQb tagging scale factor for light-flavor jet          & 2.6   & 2.6    & 2.4   \\
	  Across-year uncorrelated \PQb tagging scale factor for light-flavor jet        & 1.1   & 1.2    & 1.1   \\
	  Across-year correlated \PQb tagging scale factor for heavy-flavor jet          & 9.7   & 10.0   & 10.6  \\
	  Across-year uncorrelated \PQb tagging scale factor for heavy-flavor jet        & 8.0   & 8.3    & 9.2   \\
	  Pileup						           		 & 0.8   & 0.8    & 1.0   \\
	  Initial-state radiation				           		 & 1.0   & 1.0    & 1.0   \\
	  Trigger efficiency					           		 & 1.0   & 1.0    & 1.0   \\
	  Gluon splitting					           		 & 0.4   & 0.7    & 1.2   \\[\cmsTabSkip]

	  Total							           		 & 13.6  & 13.9   & 15.3  \\
  \hline
  \end{tabular}
  }
\label{tab:signal_systs_bin36}
\end{table}

\section{Results and interpretation} 
\label{sec:results} 

The signal is extracted via a simultaneous fit of \MJ templates to data in all bins of \Njet and \Nb, as shown in Table~\ref{tab:analysis_regions}. 
A binned maximum likelihood fit is performed using the CMS statistical analysis tool \textsc{Combine}~\cite{CMS:2024onh}. Observed and expected 95\% \CL upper limits are computed using an asymptotic approximation of the \CLs method detailed in Ref.~\cite{CLS1,CLS2,Asymptotic} with a profile likelihood ratio test statistic, in which systematic uncertainties are modeled as nuisance parameters following a frequentist approach.

As the \Njet and \Nb dependence of the QCD multijet and $\ttbar$ contributions may not be modeled perfectly by the simulation, the normalizations of the QCD multijet and $\ttbar$ contributions are treated as independent parameters and are allowed to move freely. To constrain the normalization of the QCD multijet background, we also use an $\Nlep=0$ CR with the same \Nb selection, and the requirement on \Njet shifted up by 2 (6--7, 8--9, and ${\geq}$10). The shift is motivated by the fact that events with fully hadronic $\ttbar$ decays have two more jets than those in which the $\ttbar$ pair decays semileptonically. The normalization of the QCD multijet and $\ttbar$ backgrounds in each (\Nb, \Njet) bin in the $\Nlep=1$ region moves simultaneously with the normalization of the corresponding (\Nb, $N_{\textrm{jet}}+2$) bin in the $\Nlep=0$ region. This allows the fit to constrain the QCD multijet background normalization in an (\Nb, \Njet) bin by the (\Nb, $N_{\textrm{jet}}+2$) bin in $\Nlep=0$ region where this background is dominant. On the other hand, the normalization of the $\ttbar$ background, which is dominant in the $\Nlep=1$ region, is largely constrained by the same region, with a marginal contribution from the $\Nlep=0$ region in the high-\Njet and high-\Nb bins. Mismodeling of $\Nlep$ in the QCD multijet background normalization may affect this background yields in the $\Nlep=1$ region. The size of this effect is estimated by comparing the lepton isolation and identification variables between the QCD multijet background simulation and data after subtracting all other backgrounds. An associated systematic uncertainty is assigned by allowing the relative QCD multijet normalization between the $\Nlep=0$ and $\Nlep=1$ regions to vary within its estimated uncertainty, which ranges 14--40\%, depending on the \Njet bin. For the \wjets background, a global free parameter is used for its overall normalization. To account for possible mismodeling of the \Njet distribution in MC simulation, the relative normalization between adjacent \Njet bins is allowed to vary within an uncertainty, at the level of up to 50\%, measured in the region dominated by \dyjets events.

Each main background process has dedicated $\kappa$ factors for each \Njet bin. However, all \Nb bins share the same $\kappa$ factors, so they are constrained in low \Nb bins that have larger statistical power to constrain them. The $\kappa_1$ and $\kappa_2$ factors are allowed to move independently. The $\kappa$ factors for $\ttbar$ are measured in the $\Nb=1$ region. Therefore, this region is not used to constrain them in the global fit. The fit is performed simultaneously across both CR and SR bins. The nuisance parameters are constrained and the total post-fit background uncertainties are approximately 15\% in the most sensitive region (the third \MJ bin in $\Nb\geq3$ and $\Njet\geq8$). The absence of ill-behaved nuisance parameters supports the robustness of the fit model.

\begin{figure}[htbp]
  \centering
  \includegraphics[width=\textwidth]{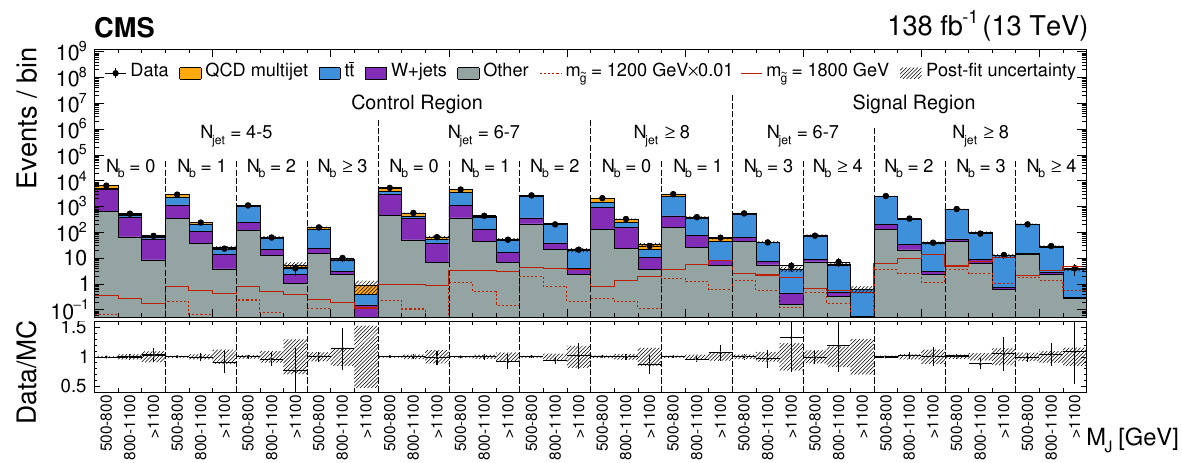}
  \caption{Background-only post-fit \MJ distributions in the $\Nlep=1$ region for all three data-taking years, with pre-fit signal overlaid. The three \MJ bins used, defined as $500<\MJ<800$, $800<\MJ<1100$, and $\MJ>1100\GeV$, are shown from left to right. The data-to-simulation ratio is shown in the lower panel, where the hatched band represents the total (statistical and systematic) uncertainty in the background prediction.}
\label{fig:postfit}
\end{figure}

\begin{figure}[htbp]
\centering
\includegraphics[width=0.75\textwidth]{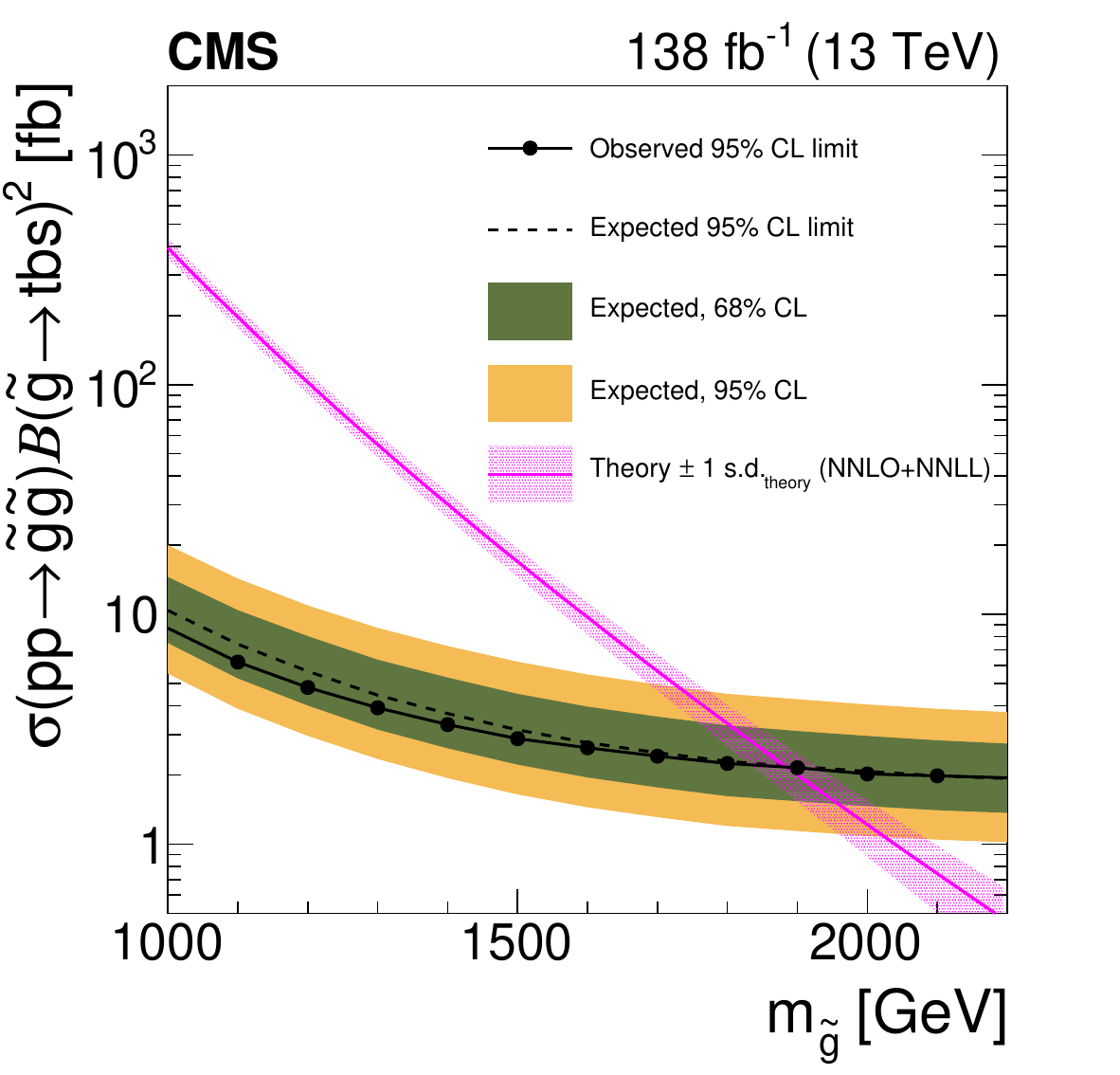}
\caption{Cross section upper limits at 95\% \CL compared to the predicted cross section of gluino pair production with $\PSg\to\PQt\PQb\PQs$ (magenta). The theoretical uncertainties in the cross section are represented by the dotted band corresponding to ${\pm}1$ standard deviation (s.d.). The expected limits are shown by the black dashed line, with 68\% and 95\% \CL variations indicated by the green inner and yellow outer bands, respectively. The observed limit is shown as a black solid line with dots.}
\label{fig:limit}
\end{figure}

Figure~\ref{fig:postfit} shows the background-only post-fit \MJ distributions in all analysis bins, while Table~\ref{tab:postfit} presents the corresponding yields in the SR. No significant excess of data over the background prediction is observed. Exclusion is set on the signal cross section. Figure~\ref{fig:limit} shows the expected and observed upper limit on cross section times branching ratio at 95\% \CL. Comparing that limit to the cross section for gluino pair production, we exclude $m_{\PSg}<1890\GeV$.

\begingroup
\renewcommand{\arraystretch}{1.1}
\begin{table}[htbp]
\centering
\topcaption{Background-only post-fit yields in the SR for all three data-taking years, with pre-fit signal yields. The uncertainty shown for total background corresponds to the post-fit uncertainty, while that for the signal is statistical only.}
\label{tab:postfit}
\vspace{1mm}
\cmsTable{%
\begin{tabular}{c c c r r r r r r r r}
\hline
$\Nb$ & $\Njet$ & $\MJ$ [GeV] & QCD multijet & $\ttbar$ & \wjets & Other & Total bkg. & Data & $m_{\PSg}=1200\GeV$ & $m_{\PSg}=1800\GeV$\\[\cmsTabSkip]
\hline
\multirow{3}{*}{$2$} &  & 500--800 & 114 & 2210 & 76 & 126 & $2520 \pm 50$ & 2516 & $353 \pm 19$ & $6.05 \pm 0.07$ \\
 & ${\ge}8$ & 800--1100 & 23 & 271 & 13.8 & 21 & $329 \pm 15$ & 336 & $259 \pm 16$ & $9.71 \pm 0.09$ \\
 &  & ${\ge}1100$ & 1.3 & 34 & 0.8 & 2.4 & $39 \pm 5$ & 39 & $114 \pm 11$ & $13.36 \pm 0.10$ \\
\cline{2-11}
\multirow{6}{*}{$3$} &  & 500--800 & 21 & 451 & 17.9 & 46 & $535 \pm 20$ & 534 & $140 \pm 12$ & $2.51 \pm 0.04$ \\
 & 6--7 & 800--1100 & 1.3 & 33 & 1.99 & 5.5 & $41.5 \pm 3.4$ & 40 & $57 \pm 8$ & $2.20 \pm 0.04$ \\
 &  & ${\ge}1100$ & 0 & 3.3 & 0.28 & 0.16 & $3.8 \pm 0.9$ & 5 & $12.4 \pm 3.5$ & $1.76 \pm 0.04$ \\
\cline{2-11}
 &  & 500--800 & 15.8 & 708 & 12.6 & 43 & $779 \pm 28$ & 790 & $351 \pm 19$ & $4.92 \pm 0.06$ \\
 & ${\ge}8$ & 800--1100 & 0.7 & 90 & 2.6 & 6.4 & $100 \pm 6$ & 88 & $255 \pm 16$ & $7.56 \pm 0.08$ \\
 &  & ${\ge}1100$ & 0 & 11.6 & 0.13 & 0.62 & $12.3 \pm 1.8$ & 13 & $106 \pm 10$ & $10.17 \pm 0.09$ \\
\cline{2-11}
 \multirow{6}{*}{${\ge}4$} &  & 500--800 & 2.3 & 63 & 1.59 & 7.1 & $74 \pm 8$ & 73 & $46 \pm 7$ & $0.670 \pm 0.022$ \\
 & 6--7 & 800--1100 & 0.7 & 4.7 & 0.132 & 0.32 & $5.9 \pm 1.0$ & 7 & $17 \pm 4$ & $0.564 \pm 0.021$ \\
 &  & ${\ge}1100$ & 0 & 0.56 & 0.034 & 0.022 & $0.62 \pm 0.19$ & 0 & $3.6 \pm 1.9$ & $0.443 \pm 0.018$ \\
\cline{2-11}
 &  & 500--800 & 3.2 & 184 & 1.58 & 14 & $203 \pm 14$ & 201 & $178 \pm 13$ & $2.10 \pm 0.04$ \\
 & ${\ge}8$ & 800--1100 & 0.5 & 24.6 & 0.39 & 2.3 & $27.8 \pm 2.8$ & 29 & $136 \pm 12$ & $3.43 \pm 0.05$ \\
 &  & ${\ge}1100$ & 0 & 3.4 & 0.02 & 0.27 & $3.7 \pm 0.6$ & 4 & $54 \pm 7$ & $4.57 \pm 0.06$ \\
\hline
\end{tabular}
}
\end{table}
\endgroup

\newpage

\section{Summary} 
\label{sec:summary} 

This analysis investigates the existence of physics beyond the standard model in events characterized by high multiplicities of jets and bottom quark (\PQb) jets in the single-lepton final state, without a requirement on missing transverse momentum. The scalar sum of the large-radius jet masses (\MJ) in an event is used for both background rejection and prediction. A simultaneous fit to the \MJ templates, in bins of jet multiplicity \Njet and the number of \PQb-tagged jets \Nb, is performed to extract potential excesses in data. The templates, derived from simulation, are corrected by measurements comparing data and simulation in dedicated control regions. This method enables a data-driven prediction of background across bins of \Njet, \Nb, and \MJ. 

The observed event yields are consistent with the background predictions. The result is interpreted within the framework of $R$-parity violating supersymmetry under the minimal flavor violating scenario. The search focuses on gluino pair production, where each gluino decays to a top, a bottom, and a strange quark via the $\lambda^{\prime\prime}_{323}$ coupling. Using proton-proton collisions at $\sqrt{s}=13\TeV$ from the CERN LHC collected by the CMS experiment from 2016 to 2018 and corresponding to an integrated luminosity of 138\fbinv, gluinos are excluded up to a mass of 1890\GeV at 95\% confidence level. This result exceeds existing limits on the considered gluino pair production model.
\begin{acknowledgments}
We congratulate our colleagues in the CERN accelerator departments for the excellent performance of the LHC and thank the technical and administrative staffs at CERN and at other CMS institutes for their contributions to the success of the CMS effort. In addition, we gratefully acknowledge the computing centers and personnel of the Worldwide LHC Computing Grid and other centers for delivering so effectively the computing infrastructure essential to our analyses. Finally, we acknowledge the enduring support for the construction and operation of the LHC, the CMS detector, and the supporting computing infrastructure provided by the following funding agencies: SC (Armenia), BMBWF and FWF (Austria); FNRS and FWO (Belgium); CNPq, CAPES, FAPERJ, FAPERGS, and FAPESP (Brazil); MES and BNSF (Bulgaria); CERN; CAS, MoST, and NSFC (China); MINCIENCIAS (Colombia); MSES and CSF (Croatia); RIF (Cyprus); SENESCYT (Ecuador); ERC PRG and PSG, TARISTU24-TK10 and MoER TK202 (Estonia); Academy of Finland, MEC, and HIP (Finland); CEA and CNRS/IN2P3 (France); SRNSF (Georgia); BMFTR, DFG, and HGF (Germany); GSRI (Greece); MATE and NKFIH (Hungary); DAE and DST (India); IPM (Iran); SFI (Ireland); INFN (Italy); MSIT and NRF (Republic of Korea); MES (Latvia); LMTLT (Lithuania); MOE and UM (Malaysia); BUAP, CINVESTAV, CONACYT, LNS, SEP, and UASLP-FAI (Mexico); MOS (Montenegro); MBIE (New Zealand); PAEC (Pakistan); MSHE, NSC, and NAWA (Poland); FCT (Portugal); MESTD (Serbia); MICIU/AEI and PCTI (Spain); MOSTR (Sri Lanka); Swiss Funding Agencies (Switzerland); MST (Taipei); MHESI (Thailand); TUBITAK and TENMAK (T\"{u}rkiye); NASU (Ukraine); STFC (United Kingdom); DOE and NSF (USA).

\begin{sloppy}
\setlength\emergencystretch{\hsize}
\hyphenation{Rachada-pisek} Individuals have received support from the Marie-Curie program and the European Research Council and Horizon 2020 Grant, contract Nos.\ 675440, 724704, 752730, 758316, 765710, 824093, 101115353, 101002207, 101001205, and COST Action CA16108 (European Union); the Leventis Foundation; the Alfred P.\ Sloan Foundation; the Alexander von Humboldt Foundation; the Science Committee, project no. 22rl-037 (Armenia); the Fonds pour la Formation \`a la Recherche dans l'Industrie et dans l'Agriculture (FRIA) and Fonds voor Wetenschappelijk Onderzoek contract No. 1228724N (Belgium); the Beijing Municipal Science \& Technology Commission, No. Z191100007219010, the Fundamental Research Funds for the Central Universities, the Ministry of Science and Technology of China under Grant No. 2023YFA1605804, the Natural Science Foundation of China under Grant No. 12535004, and USTC Research Funds of the Double First-Class Initiative No.\ YD2030002017 (China); the Ministry of Education, Youth and Sports (MEYS) of the Czech Republic; the Shota Rustaveli National Science Foundation (Georgia); the Deutsche Forschungsgemeinschaft (DFG), among others, under Germany's Excellence Strategy -- EXC 2121 ``Quantum Universe" -- 390833306, and under project number 400140256 - GRK2497; the Hellenic Foundation for Research and Innovation (HFRI), Project Number 2288 (Greece); the Hungarian Academy of Sciences, the New National Excellence Program - \'UNKP, the NKFIH research grants K 131991, K 138136, K 143460, K 143477, K 147557, K 146913, K 146914, K 147048, TKP2021-NKTA-64, and 2025-1.1.5-NEMZ\_KI-2025-00004, and MATE KKP and KKPCs Research Excellence and Flagship Research Groups grants (Hungary); the Council of Science and Industrial Research, India; ICSC -- National Research Center for High Performance Computing, Big Data and Quantum Computing, FAIR -- Future Artificial Intelligence Research, and CUP I53D23001070006 (Mission 4 Component 1), funded by the NextGenerationEU program, the Italian Ministry of University and Research (MUR) under Bando PRIN 2022 -- CUP I53C24002390006, PRIN PRIMULA 2022RBYK7T (Italy); the Latvian Council of Science; the Ministry of Science and Higher Education, project no. 2022/WK/14, and the National Science Center, contracts Opus 2021/41/B/ST2/01369, 2021/43/B/ST2/01552, 2023/49/B/ST2/03273, and the NAWA contract BPN/PPO/2021/1/00011 (Poland); the Funda\c{c}\~ao para a Ci\^encia e a Tecnologia (Portugal); the National Priorities Research Program by Qatar National Research Fund; MICIU/AEI/10.13039/501100011033, ERDF/EU, ``European Union NextGenerationEU/PRTR", projects PID2022-142604OB-C21, PID2022-139519OB-C21, PID2023-147706NB-I00, PID2023-148896NB-I00, PID2023-146983NB-I00, PID2023-147115NB-I00, PID2023-148418NB-C41, PID2023-148418NB-C42, PID2023-148418NB-C43, PID2023-148418NB-C44, PID2024-158190NB-C22, RYC2021-033305-I, RYC2024-048719-I, CNS2023-144781, CNS2024-154769 and Plan de Ciencia, Tecnolog{\'i}a e Innovaci{\'o}n de Asturias, Spain; the Chulalongkorn Academic into Its 2nd Century Project Advancement Project, the National Science, Research and Innovation Fund program IND\_FF\_68\_369\_2300\_097, and the Program Management Unit for Human Resources \& Institutional Development, Research and Innovation, grant B39G680009 (Thailand); the Eric \& Wendy Schmidt Fund for Strategic Innovation through the CERN Next Generation Triggers project under grant agreement number SIF-2023-004; the Kavli Foundation; the Nvidia Corporation; the SuperMicro Corporation; the Welch Foundation, contract C-1845; and the Weston Havens Foundation (USA).
\end{sloppy}
\end{acknowledgments}

\bibliography{auto_generated}
\cleardoublepage \appendix\section{The CMS Collaboration \label{app:collab}}\begin{sloppypar}\hyphenpenalty=5000\widowpenalty=500\clubpenalty=5000\cmsinstitute{Yerevan Physics Institute, Yerevan, Armenia}
{\tolerance=6000
A.~Hayrapetyan, V.~Makarenko\cmsorcid{0000-0002-8406-8605}, A.~Tumasyan\cmsAuthorMark{1}\cmsorcid{0009-0000-0684-6742}
\par}
\cmsinstitute{Institut f\"{u}r Hochenergiephysik, Vienna, Austria}
{\tolerance=6000
W.~Adam\cmsorcid{0000-0001-9099-4341}, L.~Benato\cmsorcid{0000-0001-5135-7489}, T.~Bergauer\cmsorcid{0000-0002-5786-0293}, M.~Dragicevic\cmsorcid{0000-0003-1967-6783}, P.S.~Hussain\cmsorcid{0000-0002-4825-5278}, M.~Jeitler\cmsAuthorMark{2}\cmsorcid{0000-0002-5141-9560}, N.~Krammer\cmsorcid{0000-0002-0548-0985}, A.~Li\cmsorcid{0000-0002-4547-116X}, D.~Liko\cmsorcid{0000-0002-3380-473X}, M.~Matthewman, J.~Schieck\cmsAuthorMark{2}\cmsorcid{0000-0002-1058-8093}, R.~Sch\"{o}fbeck\cmsAuthorMark{2}\cmsorcid{0000-0002-2332-8784}, M.~Shooshtari\cmsorcid{0009-0004-8882-4887}, M.~Sonawane\cmsorcid{0000-0003-0510-7010}, W.~Waltenberger\cmsorcid{0000-0002-6215-7228}, C.E.~Wulz\cmsAuthorMark{2}\cmsorcid{0000-0001-9226-5812}
% ...
\par}
$^{1}$Also at Yerevan State University, Yerevan, Armenia\\
$^{2}$Also at Technische Universit\"{a}t Wien, Vienna, Austria\\
% ...
\end{sloppypar}
\end{document}